# A non-perturbative study of gauge theory on a non-commutative plane

**Wolfgang Bietenholz and Frank Hofheinz**

*Institut für Physik, Humboldt Universität zu Berlin*
*Invalidenstr. 110, D-10115 Berlin, Germany*
*E-mail:* bietenho@physik.hu-berlin.de, hofheinz@physik.hu-berlin.de

**Jun Nishimura**

*Dept. of Physics, Nagoya University*
*Nagoya 464-8602, Japan*
*E-mail:* nisimura@eken.phys.nagoya-u.ac.jp

ABSTRACT: We perform a non-perturbative study of pure gauge theory in a two dimensional non-commutative (NC) space. On the lattice, it is equivalent to the twisted Eguchi-Kawai model, which we simulated at $N$ ranging from 25 to 515. We observe a clear large-$N$ scaling for the 1- and 2-point function of Wilson loops, as well as the 2-point function of Polyakov lines. The 2-point functions agree with a universal wave function renormalization. Based on a Morita equivalence, the large-$N$ double scaling limit corresponds to the continuum limit of NC gauge theory, so the observed large-$N$ scaling demonstrates the non-perturbative renormalizability of this NC field theory. The area law for the Wilson loops holds at small physical area as in commutative 2d planar gauge theory, but at large areas we find an oscillating behavior instead. In that regime the phase of the Wilson loop grows linearly with the area. This agrees with the Aharonov-Bohm effect in the presence of a constant magnetic field, identified with the inverse non-commutativity parameter.

# Contents

## 1. Introduction

The concept of spaces with NC coordinates has been known for a number of decades (see, e.g. [1]), before the mathematicians worked out a precise formulation of field theory on such geometries [2]. In the recent years, these NC field theories experienced a tremendous boom, which reflects a wide-spread belief that this variation of standard (commutative) field theory is well motivated. The main reason for this popularity is that they are obtained as low energy limits from string theories and M-theory with antisymmetric tensor fields [3, 4, 5]. Moreover, there may be practical applications to the quantum Hall effect [6]. For recent reviews we refer to refs. [7].

As an important deviation from standard field theory, the non-commutativity tensor

$$\Theta_{\mu\nu} = i\,[\hat{x}_\nu, \hat{x}_\mu] \qquad (\hat{x}_\mu : \text{ hermitean operators})$$
$$= \theta\,\epsilon_{\mu\nu} \qquad (\text{in } d=2) \tag{1.1}$$

introduces some *non-locality* inside a short range, which is characterized by the scale of the parameter $\theta$. This non-locality implies deep conceptual problems, but on the other hand it is considered a virtue which could be crucial for the link to quantum gravity as well as string and M-theory. Moreover, there was originally some hope that this property could help to weaken or even remove the UV divergences [1].

However, it is exactly the question of renormalizability where NC field theory runs into problems which are even worse than in standard field theory. UV singularities are still present, and in addition new IR singularities are mixed in (this is discussed for instance in ref. [8]). In perturbation theory, this notorious UV/IR mixing affects certain non-planar diagrams [9]. Intuitively one may grasp from eq. (1.1) that short distance effects in one direction translate into long range effects in the other directions and vice versa, which causes this kind of mixed singularities in non-planar diagrams, which do not occur in standard field theory. This property poses severe difficulties for the renormalization of perturbation theory, which are not overcome yet. For instance, scalar fields can become unstable (tachyons) due to IR effects as the non-commutativity is switched on [9]. The attempts to renormalize perturbation theory include the methods known from standard field theory (see e.g. ref. [10]), the application of Wilson's renormalization group technique [11], controlling IR divergences in the framework of supersymmetry [12] and the Hartree method [13]. In spite of some plausibility arguments in favor of perturbative renormalizability, it is an open question if NC quantum field theories do really have finite UV and IR limits. There are even claims that basic NC field theories, like NC QED, are not renormalizable [14].

Our approach is the non-perturbative study of NC field theory by means of Monte Carlo simulations. In this paper we present — as a pilot project — numeric results for 2d NC gauge theory.[1] In this particular model, the UV/IR mixing does not occur *perturbatively* since there are no UV divergent diagrams in the commutative case. However, such mixing effects are still possible on the non-perturbative level, and indeed our study is going to reveal their existence. For interesting analytical work in this field, see refs. [16]. In particular, Paniak and Szabo worked out an explicit form for the corresponding partition function. Due to Morita equivalence, 2d NC U($n$) gauge theory can be identified with the twisted Eguchi-Kawai model (TEK) [17] in the large-$N$ limit [18]. The TEK at finite $N$ corresponds to lattice NC gauge theory [19], as we briefly review in section 2. In section 3 we present our simulation results for the TEK, which do reveal that 2d NC gauge theory is renormalizable. We also observe the similarities and differences from commutative gauge theory and discuss their interpretation. Our conclusions are given in section 4.

## 2. The twisted Eguchi-Kawai model as a formulation of NC gauge theory

Back in 1982, Eguchi and Kawai conjectured that standard U($N$) and SU($N$) lattice gauge theories may be equivalent to their dimensional reductions to $d = 0$ (one point) in the large-$N$ limit [20]. Then all link variables are replaced by $U_{x,\mu} \to U_\mu$, and the plaquette action simplifies to

$$S_{EK}[U] = -N\beta \sum_{\mu \neq \nu} \text{Tr}\left(U_\mu U_\nu U_\mu^\dagger U_\nu^\dagger\right), \tag{2.1}$$

where $\mu, \nu = 1, \ldots, d$. Eguchi and Kawai proved that the Schwinger-Dyson equations — and therefore the Wilson loops — are unaltered under dimensional reduction, assuming that the U(1)$^d$ symmetry of the phases (or center symmetry) is not spontaneously broken. This is correct in $d = 2$ of course, but in $d > 2$ it only holds at strong coupling [21], so the proof for the equivalence to the reduced model (EK equivalence) has only a limited application. However, it was shown recently that a variant of the 4d Eguchi-Kawai model (EK model) surprisingly obeys the area law in a significant range of scale, so EK equivalence may still be in business in some finite regime [22].

In order to avoid the problem with the limited viability of the proof for EK equivalence, the EK model was soon modified such that the phase symmetries are preserved. This was achieved by the "quenched EK model" [21], and in even dimensions also by the TEK [17],

$$S_{\text{TEK}}[U] = -N\beta \sum_{\mu \neq \nu} \mathcal{Z}_{\mu\nu} \, \text{Tr}\left(U_\mu U_\nu U_\mu^\dagger U_\nu^\dagger\right). \tag{2.2}$$

---

[1]While our simulations were running, a related project was suggested by S. Profumo [15].

The factor $\mathcal{Z}_{\mu\nu}$ represents the *twist*,

$$\mathcal{Z}_{\mu\nu} = \mathcal{Z}_{\nu\mu}^* = \exp\left(\frac{2\pi i k}{L}\right), \qquad (\mu < \nu), \tag{2.3}$$

where $k$ is an integer and $L = N^{2/d}$ is the size of the system. The usual formulation sets $k = 1$. Further variants of the TEK have been invented [23, 24]. For numerical studies in $d > 2$, see refs. [17, 25] ($d = 4$) and ref. [26] ($d = 6$).

Here we are concerned with the 2d TEK. It seemed for a long time that adding a twist is not highly motivated in $d = 2$, since there even the ordinary EK model coincides with lattice gauge theory in the planar limit, which was solved analytically by Gross and Witten [27]. This agreement was shown already in the famous ref. [20], and it seemed likely to hold also with twist at $N \to \infty$. Still the 2d model was simulated with the $k = 1$ twist first in ref. [28] and later in ref. [29], where the finite $N$ effects were studied; but they did still not provide much motivation for including the twist.[2]

However the situation changed suddenly due to a new interpretation of the TEK as an equivalent description of NC gauge theory. The latter can be formulated if we multiply the fields by star products,

$$S = \frac{1}{4}\int d^2x \, \mathrm{Tr}\left(F_{\mu\nu}(x) \star F_{\mu\nu}(x)\right),$$
$$F_{\mu\nu} = \partial_\mu A_\nu - \partial_\nu A_\mu + ig(A_\mu \star A_\nu - A_\nu \star A_\mu). \tag{2.4}$$

For completeness we mention that the star product is defined as

$$f(x) \star g(x) = \left.\exp\left(\frac{i}{2}\Theta_{\mu\nu}\frac{\partial}{\partial x_\mu}\frac{\partial}{\partial y_\nu}\right)f(x)g(y)\right|_{x=y}, \tag{2.5}$$

and that under a star gauge transformation

$$A_\mu(x) \to V(x) \star A_\mu(x) \star V(x)^\dagger - \frac{i}{g}V(x) \star \frac{\partial}{\partial x_\mu}V(x)^\dagger,$$
$$V(x) \star V(x)^\dagger = 1, \tag{2.6}$$

the lagrangian $\mathrm{Tr}\,F_{\mu\nu}\star F_{\mu\nu}$ is not invariant, but the action in eq. (2.4) is star gauge invariant.

This new interpretation of twisted reduced models holds in the stringent sense of *"Morita equivalence"*. Roughly speaking, this means that the algebras involved in two models can be mapped onto each other one-to-one, in a way which maps each module and its structure properly from one algebra to the other.[3]

This kind of equivalence was found between NC gauge theory and twisted large-$N$ reduced models [18]. That relation was established by embedding the (dynamically generated) coordinates and momenta of the reduced model into matrices. These matrices can be mapped on functions, where the trace turns into an integral and the star products arise, so that one arrives at action (2.4). The construction of NC U($n$) gauge theories (for certain $n \in \{1, 2, \ldots\}$) works out in this way at $N = \infty$.[4] Ref. [33] pointed out that this transition corresponds to a Morita equivalence. At finite $N$ the conditions cannot be matched at the boundaries.

---
[2]A very recent reconsideration of the corresponding EK equivalence is presented in ref. [30].

[3]As an even stronger relation, "gauge Morita equivalence" implies that also the connections between two modules are mapped onto each other [31].

However, in refs. [19] this mapping was refined and it turned out that certain reduced twisted U($N$) models at finite $N$ correspond to NC Yang-Mills theories on the lattice. The corresponding lattice formulation itself was also elaborated in refs. [19].[5] From the UV/IR mixing it can be understood that such a formulation appears necessarily on a periodic lattice, so that the singularities are consistently regularized, and also the non-commutativity parameter $\theta$ takes discrete values. The continuum limit of lattice NC gauge theory coincides precisely with the large-$N$ limit of the reduced (twisted, commutative) model. This observation completes the explicit non-perturbative ("constructive") formulation of NC gauge theory, in a form which can be studied by numerical simulations. This would be very difficult in the direct lattice formulation, due to the problems related to the star product: first it is complicated to introduce star unitary link variables and a measure for them, then one would still have to find a way to compute a discrete version of the star products in action (2.4).

The U($N$) symmetry of the twisted reduced models corresponds to the gauge invariance (2.6) in the NC gauge theories. In the twisted reduced model the formulation of U($N$) invariant observables (Wilson loops, in particular) is straightforward, and the mapping on NC gauge theories led also to the formulation of NC Wilson loops [32]. This formulation works perfectly on the lattice as well [19]. Various aspects of gauge invariant observables have been investigated in refs. [35].

In order to remain in the framework of the *discrete Morita equivalence* of refs. [19] one has to take the large-$N$ limit from specific finite $N$ TEK formulations. For the 2d TEK this prescription is different from ref. [29], which used $k = 1$ and $N$ even. For the equivalence to lattice NC gauge theory with rank-one gauge group, the twist in eq. (2.2) has to be taken as

$$\mathcal{Z}_{12} = \mathcal{Z}_{21}^* = \exp\left(\pi i \frac{(L+1)}{L}\right), \tag{2.7}$$

which implies $2k = L + 1$, hence $L$ must be *odd* (and $L = N$ in two dimensions). Then there exists an explicit map from matrices onto NC fields [36]. Corresponding to $N \times N$ matrices, one obtains fields on a NC torus, which is discretized into a $N \times N$ periodic lattice. The number of degrees of freedom matches exactly. The non-commutativity is encoded in the discretized version of the star product, which is used to define products of fields in the action. If we introduce a lattice spacing $a$, the (dimensionful) non-commutativity parameter takes the form

$$\theta = \frac{1}{\pi} N a^2. \tag{2.8}$$

In the context of EK equivalence, one takes the limit $N \to \infty$ at fixed $\beta$, which ensures the equivalence of the TEK model to U($\infty$) commutative lattice gauge theory. This is called the *planar* large-$N$ limit, since only planar diagrams survive. Then the continuum limit should be taken by sending $a \to 0$ along with $\beta \to \infty$ in a particular way dictated by the coupling constant renormalization. In the present $d = 2$ case, the coupling constant scaling is purely canonical (see below)

$$\beta \propto a^{-2}. \tag{2.9}$$

---

[4]On the other hand, there are severe problems in formulating SU($n$) gauge theory on a NC space.
[5]For earlier work on a lattice formulation of NC gauge theory, see e.g. ref. [34].

Now interpreting the TEK as a lattice NC gauge theory, the planar limit corresponds to $\theta = \infty$ due to eq. (2.8). This is a non-perturbative account of the well-known equivalence between $\theta = \infty$ NC field theory and the corresponding (commutative) planar field theory.

In this paper, we are interested in a continuum limit of the lattice NC gauge theory at *finite $\theta$*. According to eq. (2.8), this requires the large-$N$ limit to be taken simultaneously with the continuum limit $a \to 0$.[6] This also implies that non-planar diagrams are included, but in this specific model they do not involve perturbative UV/IR mixing, as we mentioned in the introduction. As we are going to explain in the next section, $\beta$ has to be scaled as a function of the lattice spacing $a$ in exactly the same way as in the planar theory. Hence in $d = 2$ we are going to search for a *double scaling limit* keeping the ratio $N/\beta$ constant. The question of renormalizability of 2d NC gauge theory at finite $\theta$ can be answered by studying this large-$N$ limit of the TEK. If the observables converge to finite values, we can conclude that 2d lattice NC gauge theory does have a finite continuum limit. Our results will also be presented in the next section.

A double scaling limit of the *untwisted* EK model was searched for in ref. [29]. Large-$N$ scaling of observables was indeed observed at a fixed ratio $N/\beta$, but this combination of parameters was identified empirically. The TEK, on the other hand, was studied only at fixed $\beta$ in ref. [29] and did not reveal systematic finite $N$ effects. However, we repeat that the TEK was studied there with twist parameter $k = 1$ and even values for $N$.

## 3. Numeric results

In this section we present our numeric simulation results for the 2d TEK at odd $N$, with the twist specified in eq. (2.7). In the form of eq. (2.2), the action cannot be simulated with a heat bath algorithm because it is not linear in the matrices $U_\mu$. Following ref. [37] we introduce an auxiliary matrix field $Q$, which enters in a gaussian form and which linearizes the action in the $U_\mu$,

$$S_{\text{TEK}}[U, Q] = N\beta \sum_{\mu < \nu} \text{Tr}\Big[Q^\dagger_{\mu\nu} Q_{\mu\nu} - Q^\dagger_{\mu\nu}\left(t_{\mu\nu} U_\mu U_\nu + t_{\nu\mu} U_\nu U_\mu\right) - \\ - \left(t^*_{\mu\nu} U^\dagger_\mu U^\dagger_\nu + t^*_{\nu\mu} U^\dagger_\nu U^\dagger_\mu\right) Q_{\mu\nu}\Big]. \tag{3.1}$$

The auxiliary field $Q$ consists of general complex $N \times N$ matrices, and their integration reproduces the TEK as given in eq. (2.2), if $t_{\mu\nu} = \sqrt{\mathcal{Z}_{\mu\nu}}$. In this respect, and in all other technical points of the simulation, we follow the methods which were used and described

---

[6]The physical extent of the NC torus $Na$ goes to infinity in this limit. In order to obtain a finite torus in the continuum limit, one would have to use a different formulation of lattice NC gauge theory [19].

– 5 –

| $N$ | 25 | 35 | 55 | 85 | 125 | 195 | 255 | 515 |
|---|---|---|---|---|---|---|---|---|
| # of configurations | 9000 | 8400 | 20000 | 8000 | 3400 | 260 | 120 | 18 |

**Table 1:** Overview of the statistics in our simulation of the TEK.

in ref. [29]. For instance, we also update SU(2) subgroups of the $U_\mu$ matrices, in the spirit of Cabbibo and Marinari [38].

Table 1 displays the number of configurations that we generated at various $N$. Note that the few configurations at our largest values of $N$ provide conclusive signals only for part of the observables that we evaluated, see below. In most of our study we fixed

$$\frac{N}{\beta} = 32, \qquad (3.2)$$

which corresponds to

$$\theta = \frac{1}{4\pi} \frac{N}{\beta} = \frac{8}{\pi} \approx 2.55 \qquad (3.3)$$

in the continuum limit.

As a point of orientation for the interpretation of our data, we consider the exact solution of planar U($\infty$) and SU($\infty$) lattice gauge theory in $d = 2$ [27]. There the Wilson loop of a rectangle $I \times J$ follows the area law

$$w(I \times J) = \exp(-\kappa(\beta)IJ), \qquad (3.4)$$

where the (dimensionless) string tension $\kappa(\beta)$ is given by

$$\kappa(\beta) = \begin{cases} -\ln \beta & \beta \leq \frac{1}{2} \\ -\ln\left(1 - \frac{1}{4\beta}\right) & \beta \geq \frac{1}{2} \end{cases}. \qquad (3.5)$$

In particular, the internal energy amounts to

$$E(\beta) = w(1 \times 1) = \begin{cases} \beta & \beta \leq \frac{1}{2} \\ 1 - \frac{1}{4\beta} & \beta \geq \frac{1}{2} \end{cases}, \qquad (3.6)$$

which reveals a third order phase transition.[7] Eq. (3.4) shows how one should tune the bare coupling constant $\beta$ as a function of $a$ when one takes the continuum limit $a \to 0$. In fact in this case the scaling is exact if one takes the lattice spacing $a$ to be

$$a = \sqrt{\kappa(\beta)}. \qquad (3.7)$$

Then the Wilson loop becomes an $a$-independent function of the physical area $a^2 IJ$. From eq. (3.7) we see that $\beta$ should be sent to infinity in the continuum limit as in relation (2.9).

The EK equivalence implies that the exact result (3.4) is reproduced by considering the corresponding Wilson loops in the TEK in the large-$N$ limit at fixed $\beta$. More specifically, we define 'Wilson loops' in the TEK model as [17]

$$W_{\mu\nu}(I \times J) = (\mathcal{Z}_{\mu\nu})^{I \cdot J} \, \mathrm{Tr}\left(U_\mu^I U_\nu^J U_\mu^{\dagger I} U_\nu^{\dagger J}\right), \qquad (3.8)$$

---
[7]Note that we are always in the $\beta \geq 1/2$ regime for the values of $N$ that we are going to consider.

where $I, J = 1, 2, \ldots$. For $I = J = 1$ it reduces to the plaquette, which appears in the action (2.2). Then the EK equivalence states that

$$\lim_{N \to \infty} \frac{1}{N} \langle W_{12}(I \times J) \rangle = w(I \times J), \qquad (3.9)$$

at fixed $\beta$. Here the term $w(I \times J)$ is the expectation value of Wilson loop in the U($\infty$) lattice gauge theory, given by eq. (3.4).

Let us now consider Wilson loops in NC gauge theory. Since the Wilson loop (3.8) in the TEK model is U($N$) invariant, it defines a star gauge invariant observable upon mapping the matrices onto NC fields [32]. This is precisely what we consider as the Wilson loop in NC gauge theory on the lattice [19]. Now we take the large-$N$ limit together with the continuum limit $a \to 0$ in order to obtain a finite non-commutativity parameter $\theta$ in eq. (2.8). We define a "lattice spacing" as in the planar theory, i.e. as in eq. (3.7). Then the Wilson loop still agrees with the planar result (3.4) when the physical extent of the Wilson loop is much smaller than the non-commutativity scale $\sqrt{\theta}$. (In the limit $\theta \to \infty$ it agrees with the planar result for arbitrarily large area, as we mentioned before.) Therefore, the scaling in the small area regime requires the lattice spacing to be chosen in the same way as in the planar theory. The first interesting question is now whether the scaling of the Wilson loop extends also to larger areas.

In the EK model it was observed that square shaped Wilson loops converge faster to the known exact results in the planar limit than other rectangles with the same area [29]. Hence we also focus on square shaped Wilson loops here, which corresponds to the $I = J$ case in definition (3.8). We define the normalized Wilson loop

$$W(I) = \frac{1}{N} \langle W_{12}(I \times I) \rangle, \qquad (3.10)$$

which should agree with $w(I \times I)$ in eq. (3.4) when $aI \ll \sqrt{\theta}$. Due to the presence of the twist, there is no invariance under the parity transformation $U_1 \to U_2$, $U_2 \to U_1$. As a consequence, $W_{\mu\nu}$ is complex in general, and $W_{12} = W_{21}^*$, hence the Wilson loop depends on its orientation. The real part represents an average over both orientations. Figure 1 shows the real part Re[$W(I)$] as a function of the physical loop area.

Large-$N$ scaling is clearly confirmed. At small up to moderate areas, the Wilson loop follows the area law of planar lattice gauge theory given in eq. (3.4), but at large areas it deviates and the real part oscillates around zero instead. This observation is consistent over a wide range of $N$. Remarkably, not even the absolute value decays monotonously; at large areas it seems to fluctuate around an approximately constant value. This is illustrated in figure 2 (on top).

Figure 2 (below) shows that beyond the Gross-Witten regime, the phase increases *linearly* (in the physical area) in the large-$N$ limit. Additional measurements at $N/\beta = 16$, 24 and 48, corresponding to different values of $\theta$, reveal that the phase $\Phi$ of the Wilson loop $W(I)$ (beyond the Gross-Witten regime) is, in fact, given to a very high precision by the simple relation

$$\Phi = \frac{A}{\theta}, \qquad (3.11)$$

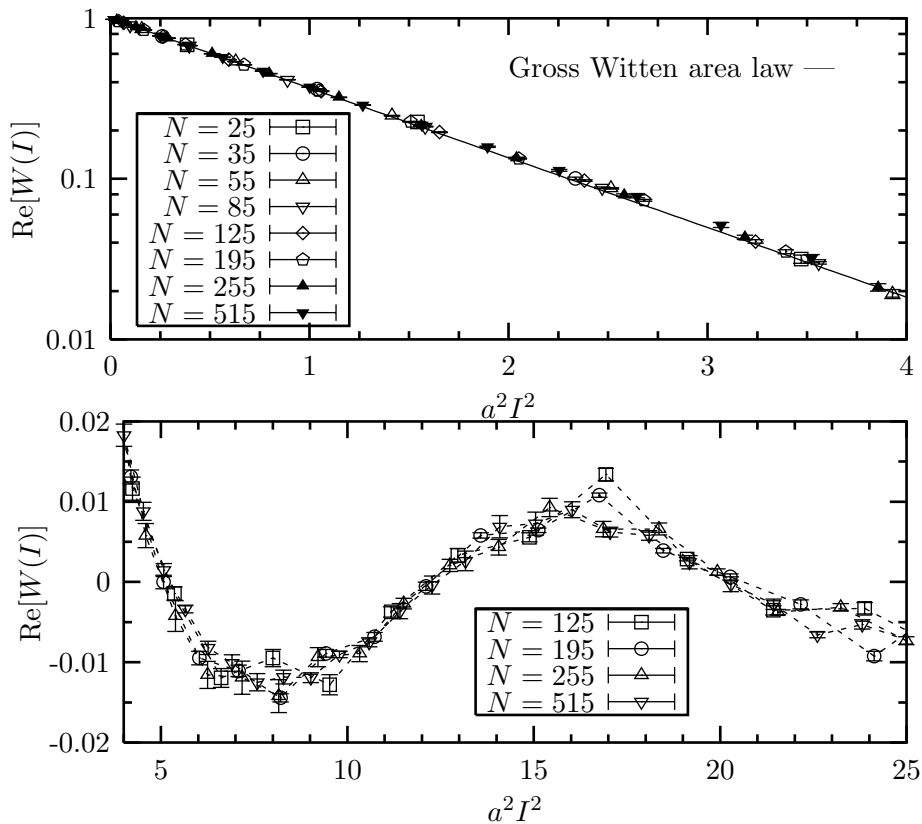

**Figure 1:** The real part of the normalized Wilson loop $W(I)$ against the physical area for various values of $N$. At small areas $\mathrm{Re}[W(I)]$ follows the Gross-Witten area law, but at larger areas it oscillates around zero. The dashed lines are plotted to guide the eye.

where $A$ is the physical loop area, see figure 3. This has been confirmed also for other (non-square) rectangular Wilson loops, which shows that the effect does not depend on the shape of the Wilson loop.[8] Indeed the formula (3.11) agrees with the Aharonov-Bohm effect in the presence of a constant magnetic field $B = 1/\theta$ across the plane. This is reminiscent of the description of NC gauge theory by Seiberg and Witten [5]. Our numeric results seem to support a picture of that kind, which is also known in solid state physics [6].

The observed large area behavior of the Wilson loop confirms that the continuum limit of NC gauge theory is different from any *ordinary* (commutative) gauge theory, hence we have found a new universality class. Naively one might think that in the large area regime $A \gg \sqrt{\theta}$, the effect of non-commutativity is invisible. The fact that we do observe the contrary can be understood as a manifestation of non-perturbative UV/IR mixing. This observation also implies that the limit $\theta \to 0$ does not commute with the continuum limit.

In the double scaling limit of the untwisted EK model, the expectation value of the Wilson loop is real, and it remains positive even at large physical areas [29]. This means that the two models, twisted and untwisted, yield qualitatively *different double scaling limits*, although they become identical in the planar large-$N$ limit.

---

[8]Note, however, that the expectation values of Wilson loops with the same area but with different shapes have in general different absolute values in the double scaling limit.

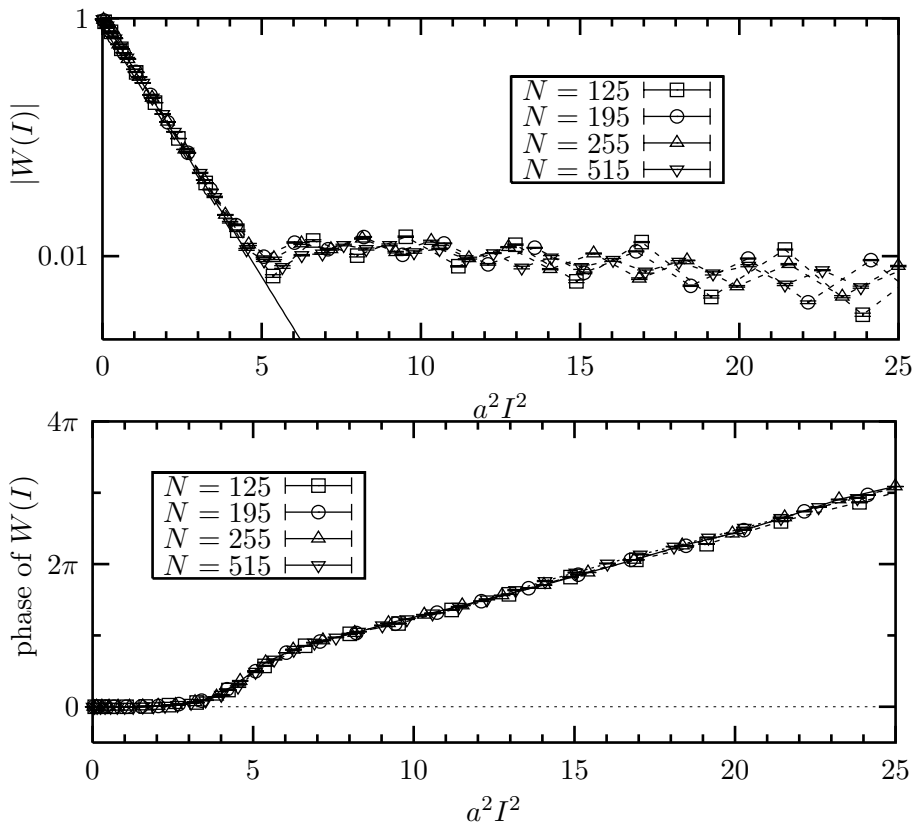

**Figure 2:** The polar coordinates of the complex Wilson loop $W(I)$ plotted against the physical area $A = a^2 I^2$. At small areas it is approximately real and follows the Gross-Witten area law (solid line), c.f. figure 1. At larger areas the absolute value does not seem to decay any more, and the phase agrees with the formula $\Phi = A/\theta$, c.f. figure 3.

Figure 4 shows the connected Wilson loop 2-point function

$$G_2^{(W)}(I) = \langle W_{12}(I \times I) W_{21}(I \times I) \rangle - \langle W_{12}(I \times I) \rangle \langle W_{21}(I \times I) \rangle \in \mathbb{R}, \tag{3.12}$$

again plotted against the physical area, for $I = 1, \ldots, N$. Here we include a wave function renormalization factor

$$G_2^{(W)} \to \beta^{-0.6} G_2^{(W)}. \tag{3.13}$$

The exponent $-0.6$ was found to be optimal for $G_2^{(W)}$ to scale. Indeed it leads to a neat large-$N$ scaling in some regime.

Next we consider the Polyakov line

$$P_\mu(I) = \text{Tr}\left(U_\mu^I\right), \tag{3.14}$$

which is also U($N$) invariant and therefore has an interpretation as a star gauge invariant observable in NC gauge theory. In fact Polyakov lines are mapped to *open* Wilson lines which carry non-zero total momentum [32]. Their momentum $\vec{p}$ is related to the separation vector $\vec{v}$ between the two ends of the line. In general the relation is given by $v_\mu = \Theta_{\mu\nu} p_\nu$ modulo the periodicity of the torus. In the present case, the Polyakov line $P_\mu(I)$ corresponds to a momentum mode with $p_\nu = 2\pi\ell/(Na)$, where the integer $\ell$ is given by $I/2$ and by $(I+N)/2$ for even and odd $I$, respectively. In what follows, we plot the results against $aI$ for even $I = 2, \ldots, N-1$.

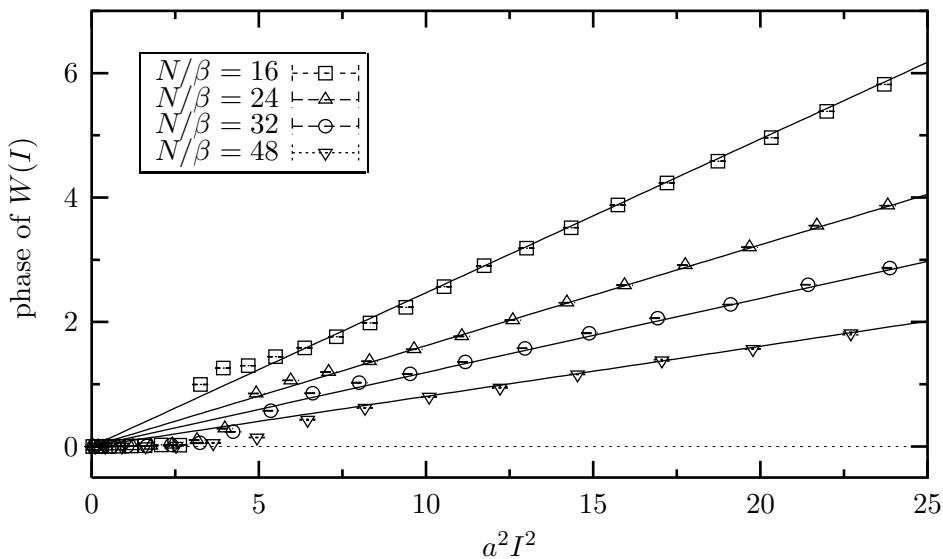

**Figure 3:** The phase of the Wilson loop $W(I)$ is plotted as a function of the physical area $A = a^2 I^2$ for $N = 125$ and various values of $N/\beta$, corresponding to $\theta = 1.27, 1.91, 2.55, 3.82$. For large areas ($A \gtrsim O(\theta)$), the phase $\Phi$ agrees with the formula $\Phi = A/\theta$, represented by the straight lines.

The phase symmetry[9] makes $\langle P_\mu(I) \rangle$ vanish, but the connected $n$-point functions ($n > 1$) of Polyakov lines are sensible observables. In figure 5 we show the 2-point function[10]

$$G_2^{(P)}(I) = \langle P_\mu(I) P_{-\mu}(I) \rangle, \qquad (3.15)$$

where $P_{-\mu}(I) = \mathrm{Tr}(U_\mu^{\dagger\,I})$. Note that there is no disconnected part in $G_2^{(P)}$. Again we insert the wave function renormalization which was optimal for the Wilson 2-point function,

$$G_2^{(P)} \to \beta^{-0.6} G_2^{(P)}. \qquad (3.16)$$

As a function of the physical length $aI$, the result is consistent with large-$N$ scaling, as well as a *universal* wave function renormalization. A similar wave function renormalization was also observed in the EK model [29], where the optimal factor in relation (3.16) is modified to $\beta^{-0.65}$.

Our statistics was not sufficient to extract conclusive signals for the connected parts of 3-point functions or even 4-point functions. For completeness we present nevertheless

---

[9] In the terminology of NC gauge theory, this corresponds to momentum conservation.

[10] The choice of the direction $\mu$ is irrelevant. In practice we average over both possibilities in order to enhance the statistics.

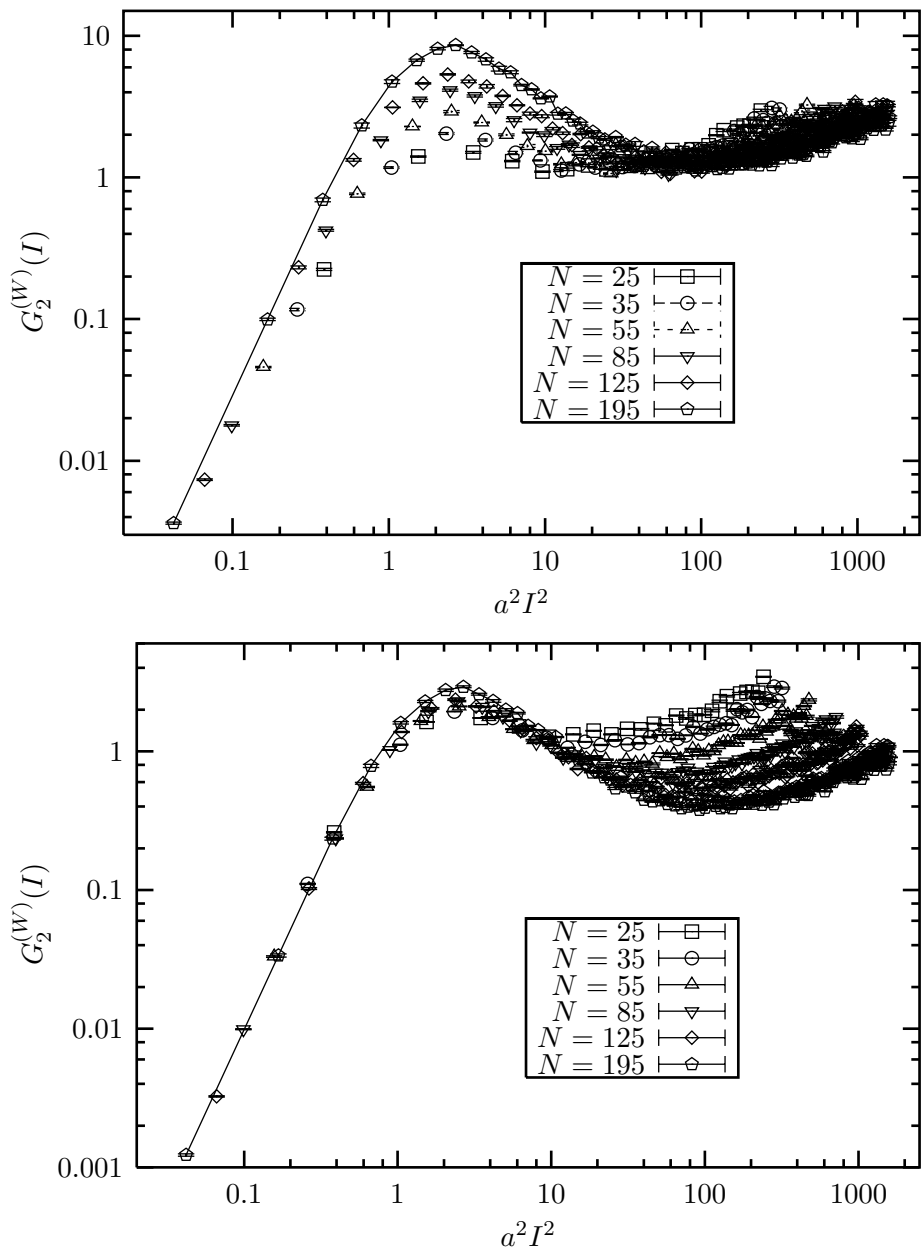

**Figure 4:** The 2-point function $G_2^{(W)}(I)$, defined in eq. (3.12), with the renormalization factor given in eq. (3.13) (bottom) and without (top), plotted against the physical area, for $I = 1, \ldots, N$. The solid line connects the results for $N = 195$.

our result for

$$G_3^{(P)}(I) = \langle P_\mu(I) P_\mu(I) P_{-\mu}(2I) \rangle \tag{3.17}$$

(there is still no disconnected part). In figure 6 we show our data after re-scaling, if we assume a universal wave function renormalization, which is not ruled out. This means that any $n$-point function is renormalized as $G_n \to \beta^{-0.6n/2} G_n$ (for $n \geq 2$).

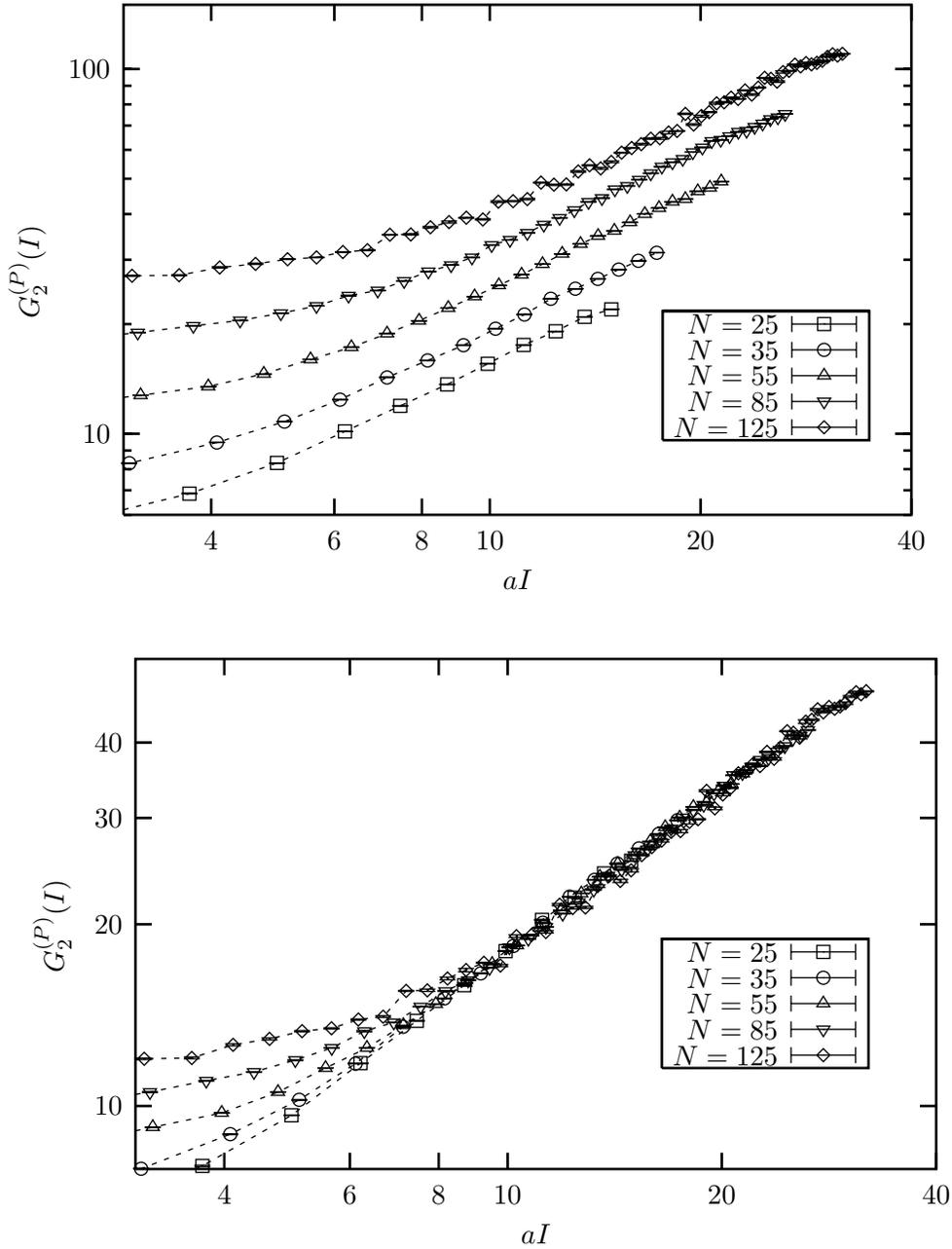

**Figure 5:** The 2-point function $G_2^{(P)}(I)$, defined in eq. (3.15), with the renormalization factor given in eq. (3.13) (bottom) and without (top), plotted against the physical distance, for $I = 2, \ldots, N-1$.

## 4. Conclusions

We have investigated the 2d TEK in the form which allows for a mapping onto 2d NC gauge theory on the lattice, by discrete Morita equivalence. In this representation as a twisted reduced model, NC gauge theory takes a form which can be simulated numerically. It requires the specific twist given in eq. (2.7) and odd values of $N$.

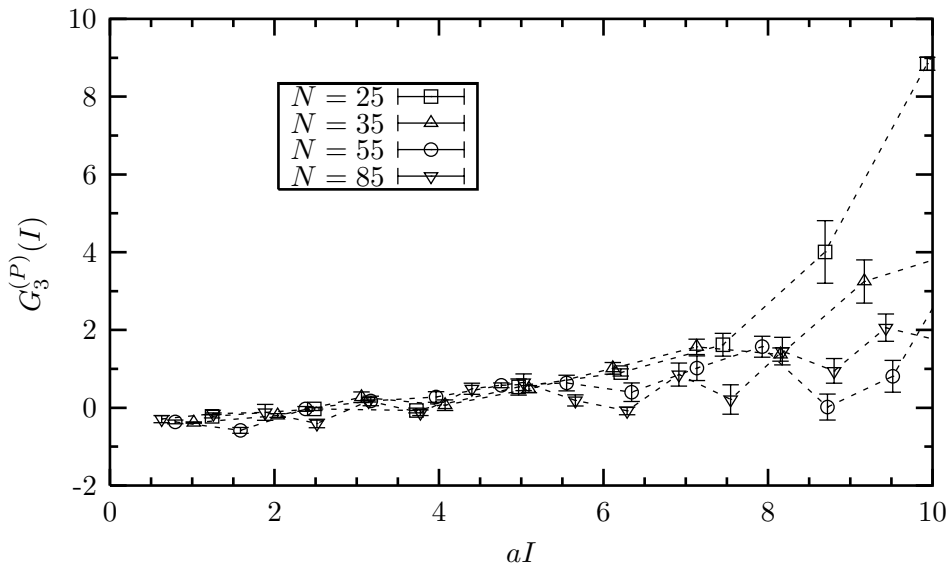

**Figure 6:** The 3-point function $G_3^{(P)}(I)$, defined in eq. (3.17) and rescaled by a universal renormalization factor, plotted against the physical distance.

The first conclusion from our simulation results is that we do observe a double scaling limit as $N, \beta \to \infty$. This corresponds to the continuum limit of the NC rank-one gauge theory, which has therefore also finite observables. This observation demonstrates the non-perturbative renormalizability of 2d NC rank-one gauge theory.

The Wilson loop follows an area law at small physical areas, and in this regime NC gauge theory agrees with planar standard gauge theory. However, at larger areas the Wilson loop becomes complex and the real part (the mean values over both loop orientations) begins to oscillate around zero. The phase is proportional to the physical area enclosed by the Wilson loop, irrespectively of its shape, and the coefficient of proportionality is given by the inverse of the non-commutativity parameter $\theta$. This agrees with the Aharonov-Bohm effect in the presence of a constant magnetic background field

$$B = \frac{1}{\theta}. \tag{4.1}$$

Our results support this law, which is also a key element of the Seiberg and Witten description of NC gauge theory [5]. Moreover, the same law also occurs in condensed matter physics: if one assumes a plane crossed by a constant magnetic flux, then the electrons in this plane can be projected to the lowest Landau level in a NC space, where $\theta = \hbar c/eB$ [6].

The behavior of the Wilson loop at large areas implies that we have found a qualitatively new universality class. At first sight, it may look surprising that the non-commutativity — which introduces a short-ranged non-locality in the action — changes the IR behavior of the gauge theory completely. However, this effect does not appear unnatural in the presence of UV/IR mixing. It is remarkable that this mixing effect does not occur in the perturbative expansion of this model, hence our results for large Wilson loops reveal a purely non-perturbative UV/IR mixing.

For the connected Wilson loop 2-point function, as well as the 2-point function of the Polyakov line, we can confirm the large-$N$ scaling. These observables are in agreement with a universal wave function renormalization, which yields a factor $\beta^{-0.6}$ for a connected 2-point function. This is similar to — but not identical with — the universal renormalization factor which was found for the (untwisted) EK model.[11]

Finally we should mention that in $d > 2$, NC field theories are often formulated such that the time coordinate is *not* included in the non-commutativity. Including it causes especially severe conceptual problems regarding causality etc. Therefore it would be important to confirm non-perturbative renormalizability also in higher dimensions, where non-commutativity is introduced only between spatial coordinates. We are currently simulating the 3d NC $\phi^4$ model [40], which can also be mapped onto a twisted reduced matrix model [32]. A self consistent one-loop analysis in $d = 4$ suggests the appearance of an interesting phase diagram [13].

## Acknowledgments

It is a pleasure to thank Harald Dorn, Philippe de Forcrand and Richard Szabo for inspiring remarks.

---

[11]On the other hand, in the hermitean matrix version of the (untwisted) reduced model, the universal wave function renormalization holds only in the presence of supersymmetry [39].